\documentclass{article}
\usepackage{latexsym}
\usepackage{amsmath}
\usepackage{amsthm}
\usepackage{amsfonts}
\allowdisplaybreaks
\title{Rotating frames and gauge invariance in two-dimensional
  many-body quantum systems}
\author{Jos\'e M\'endez Gamboa and Antonio O.\ Bouzas \thanks{E-mail:
    abouzas@mda.cinvestav.mx}\\\small Departamento de F\'{\i}sica
  Aplicada, CINVESTAV-IPN \\\small Carretera Antigua a Progreso Km.\ 
  6, Apdo.\ Postal 73 ``Cordemex''\\\small M\'erida 97310, Yucat\'an,
  M\'exico} 
\date{April 2, 2003}
\newcommand{\Ll}{\ensuremath{\mathcal{L}}}
\newcommand{\V}{\ensuremath{\mathcal{V}}}
\renewcommand{\H}{\ensuremath{\mathcal{H}}}
\newcommand{\mat}[1]{\ensuremath{\boldsymbol{#1}}}
\newcommand{\ver}[1]{\ensuremath{\boldsymbol{\hat{#1}}}}
\renewcommand{\S}{\ensuremath{\mathfrak{S}}}
\newcommand{\s}{\ensuremath{\mathfrak{s}}}
\newcommand{\Q}{\ensuremath{\mathfrak{Q}}}
\newcommand{\q}{\ensuremath{\mathfrak{q}}}
\newcommand{\R}{\ensuremath{\mathfrak{R}}}

\newcommand{\C}{\ensuremath{\mathfrak{C}}}
\newcommand{\G}{\ensuremath{\mathfrak{G}}}
\begin{document}
\maketitle
\begin{abstract}
  We study the quantization of many-body systems in two dimensions in
  rotating coordinate frames using a gauge invariant formulation of
  the dynamics.  We consider reference frames defined by linear and
  quadratic gauge conditions.  In both cases we discuss their Gribov
  ambiguities and commutator algebra.  We construct the momentum
  operators, inner-product and Hamiltonian in both types of gauges,
  for systems with and without translation invariance.  The analogy
  with the quantization of QED in non-covariant gauges is emphasized.
  Our results are applied to quasi-rigid systems in the Eckart frame.
\end{abstract}
\section{Introduction}
\label{sec:intro}

The problem of quantizing a many-body mechanical system in a rotating
reference frame is of interest both by itself and for its possible
applications to specific problems in, e.g., molecular and nuclear
physics.  When the underlying dynamics are rotationally symmetric,
which is the only case we consider, the coordinate transformation from
a space-fixed reference frame to a rotating one with the same origin
is a time dependent symmetry transformation.  It is thus appropriate
to formulate the theory in such a way that it is invariant under
symmetry transformations whose parameters depend on time, or gauge
transformations \cite{als}.  If the dynamics are described in terms of
a gauge-invariant action, since we know how to quantize a mechanical
system in a space-fixed coordinate frame, we can perform a gauge
transformation in order to obtain the quantum theory in a rotating
reference frame.  Gauge invariance guarantees that both theories are
physically equivalent.

Whereas formulating a quantum theory in a rotating frame is a
mathematical problem, the choice of the particular frame in which to
formulate the theory is dictated by the physics of the specific system
under consideration.  It is often the case that the relevant rotating
frame is defined implicitly, by restrictions on the trajectories of
the system in that frame.  In the gauge-invariant approach to the
quantization in rotating frames, such restrictions are incorporated
into the theory as gauge conditions.  The action is then given in
terms of degrees of freedom that are not independent, but must satisfy
certain functional relations.  This situation is familiar from the
theory of gauge fields \cite{lee}.  In quantum electrodynamics (QED),
for instance, the degrees of freedom are the components of
the vector potential $\mat{A}(t,\mat{x})$, which may be required to
satisfy such relations as $\mat{\nabla}\cdot\mat{A}=0$ (Coulomb
gauge), or $\mat{x}\cdot\mat{A}=0$ (multipolar gauge) \cite{jac}, at
all times $t$.

In this paper we study the quantization of many-body systems in two
dimensions in rotating coordinate frames, using a gauge-invariant
formulation.  We consider systems of $N$ spinless particles in the
plane, interacting through two-body central potentials.  We focus on
developing the formalism, which is a necessary previous step to
considering applications to realistic models.  By restricting
ourselves to the two-dimensional rotation group we separate the
treatment of the gauge-invariant formalism from the technical
intrincacies of non-abelian groups such as the three-dimensional
rotation group, which we will consider elsewhere.  

Our treatment closely follows the approach of \cite{lee,chr} to
Yang-Mills theories.  Previous treatments of the quantization of two-
and three-dimensional $N$-body systems in rotating frames within a
gauge invariant approach have been given in \cite{bs1,bs2,ljn} (and
references therein).  In \cite{bs2} the gauge symmetry is implemented
within the Hamiltonian BRST formalism \cite{hen}.  In \cite{ljn} a
point of view based on the shape-space theory of deformable bodies is
adopted.  Non-gauge-invariant treatments can be found in, e.g.,
\cite{vil} in the context of nuclear physics, and in \cite{gai} in
molecular physics.  Our goals are to establish a formal framework with
the most direct geometrical and physical interpretation, and to
provide careful and systematic derivations of our results.
Furthermore, the results we obtain are different from previous ones.

In the following section we describe the class of models considered
throughout the paper, and their gauge-invariant formulation.  Their
quantization in a space-fixed frame is given, and shown to be
equivalent to the original, non-gauge-invariant system.  In the
remaining sections we obtain the quantized theory in rotating frames
by means of a gauge transformation from the space-fixed frame.  In
section \ref{sec:linear} we consider rotating frames defined by linear
gauge conditions.  Such gauges are the most important ones for
practical applications.  We discuss the Gribov ambiguities \cite{grb}
of those gauges, which are important for the construction of the inner
product in Hilbert space.  The algebra of commutators is discussed in
detail, and an explicit realization of that algebra in terms of
differential operators is given.  We then use those operators to
construct the Hamiltonian.  The equivalence of the quantum theory in
these and other gauges with the theory formulated in a non-rotating
reference frame is kept manifest at every step.  We emphasize that by
describing the system from a rotating reference frame defined by
imposing restrictions on the coordinates, we are in fact introducing
orthogonal curvilinear coordinates in configuration space.

In section \ref{sec:quadratic} we go through the same steps, though
more briefly, to obtain the theory in the instantaneous principal axes
frame as an example of a gauge condition depending quadratically on
the coordinates.  Although the treatment is straightforward, the
results are technically much more complicated than in the linear case.
This fact makes the practical usefulness of this gauge condition
doubtful.  The quantization in linear gauges from section
\ref{sec:linear} is extended in section \ref{sec:cmm} to
translationally invariant systems in rotating frames with origin at
the center of mass.  These results are then applied to the case of
quasi rigid systems in section \ref{sec:eckart}, where we discuss the
Eckart gauge and recover some of the classic results of \cite{ek2}.
In section \ref{sec:final} we give our final remarks.  We try
throughout the paper to make manifest the parallel between our
approach and the quantization of QED in non-covariant gauges.  In
appendix \ref{sec:appqed} we give a brief summary of those aspects of
QED which are relevant to the analysis presented in the main body of
the paper. In appendix \ref{sec:appL} we derive some technical results
needed in the sequel.

\section{\mat{N}-particle system}
\label{sec:model}

We consider a system of $N$ particles in two dimensions interacting
through a two-body central potential, described by the Lagrangian
\begin{equation}
  \label{eq:rlag}
  \Ll = \frac{1}{2} \sum_{\alpha=1}^N m_\alpha \dot{\mat{r}}^2_\alpha -
  \sum_{\alpha < \beta =1}^N V(|\mat{r}_\alpha - \mat{r}_\beta |) -
  \sum_{\alpha=1}^N U(r_\alpha).
\end{equation}
If the one-body potential $U = 0$, \Ll\ is invariant under the group of
Euclidean motions of the plane. In this and the following sections we
consider $U \neq 0$ and focus on the abelian group of two-dimensional
rotations, deferring the discussion of translation invariance until
section \ref{sec:cmm}.

We adopt the passive point of view for coordinate transformations.
\Ll\ is invariant under time-independent rotations of the coordinate
frame.  In order to make \Ll\ invariant under changes of arbitrarily
rotating coordinate frames we apply the usual Yang-Mills construction
\cite{yml} to (\ref{eq:rlag}).  We add a new degree of freedom $\xi$
to the system, and postulate the following transformation law under
infinitesimal rotations of the coordinate frame,
\begin{equation}
  \label{eq:trans}
  \delta\mat{r}_\alpha = - \delta \theta \ver{z} \wedge \mat{r}_\alpha,
  \qquad
  \delta \xi = - \delta \dot{\theta},
\end{equation}
with $\delta \theta = \delta \theta(t)$ an arbitrary function of $t$
and $\ver{z}$ a unit vector orthogonal to the plane.  These are the
infinitesimal gauge transformations of the system.  We define the
covariant derivative $D_t \mat{r}_\alpha \equiv \dot{\mat{r}}_\alpha -
\xi \ver{z} \wedge \mat{r}_\alpha$, which transforms as a vector under
gauge transformations, $\delta (D_t \mat{r}_\alpha) = - \delta \theta
\ver{z} \wedge D_t \mat{r}_\alpha$.  Substituting time derivatives in
(\ref{eq:rlag}) 
by covariant derivatives, we obtain a Lagrangian invariant under
time-dependent rotations of the coordinate frame.  Explicitly, we
write,
\begin{equation}
  \label{eq:glag}
    \Ll  = \frac{1}{2} \sum_{\alpha=1}^N m_\alpha
    (D_t\mat{{r}}_\alpha)^2  - \V + \ell_z \xi
     = \frac{1}{2} \sum_{\alpha=1}^N m_\alpha \dot{\mat{r}}^2_\alpha
    + \frac{\xi^2}{2} \sum_{\alpha=1}^N m_\alpha r^2_\alpha - \xi
    \ver{z} \cdot \sum_{\alpha=1}^N m_\alpha \left( \mat{r}_\alpha
      \wedge \dot{\mat{r}}_\alpha \right) - \V + \ell_z \xi,
\end{equation}
where we denoted by \V\ the potential energy for brevity.  In
(\ref{eq:glag}) we added an extra term $\ell_z \xi$ to \Ll\ which fixes
the value of the angular momentum through the equation of motion for
$\xi$.  That term plays a role analogous to the source term
$j^0(\mat{x}) A^0(t,\mat{x})$ in electrodynamics\footnote{We remark
  that external source terms break gauge invariance when the gauge
  symmetry is non-abelian.  In that case, in order to preserve gauge
  invariance and, at the same time, to have a non-homogeneous Gauss
  law, we must incorporate the source into the theory as a dynamical
  degree of freedom.}, as discussed in appendix \ref{sec:appqed}.  \Ll\ 
is invariant under gauge transformations if the constant $\ell_z =0$,
and quasi-invariant otherwise, $\delta \Ll = - \ell_z \delta
\dot{\theta}$.

\Ll\ in (\ref{eq:glag}) describes the same dynamics as (\ref{eq:rlag}),
but from a coordinate frame rotating with angular velocity $-\xi$ with
respect to the laboratory frame \cite{lan}.  Notice, however, that
$\xi$ is a dynamical variable describing the coupling of the particles
to inertial forces.  The equations of motion for $\mat{r}_\alpha$ are
$m_\alpha D_tD_t\mat{r}_\alpha + \nabla_{\alpha} \V =0$ or, more
explicitly,
\begin{equation}
  \label{eq:eqm}
  m_\beta \mat{\ddot{r}}_\beta = 2 m_\beta \xi \ver{z} \wedge
  \dot{\mat{r}}_\beta + m_\beta \dot{\xi} \ver{z} \wedge \mat{r}_\beta
  + m_\beta \xi^2 \ver{z} \wedge (\mat{r}_\beta \wedge \ver{z}) -
  \nabla_{\beta} \V,
\end{equation}
where the terms corresponding to the Coriolis, azimuthal and
centrifugal forces are apparent \cite{mek}.  A consequence of
rotational symmetry is the conservation of the system's total angular
momentum
\[
L_z = \ver{z} \cdot\sum_{\alpha=1}^N m_\alpha (\mat{r}_\alpha \wedge
      D_t \mat{r}_\alpha).
\]
The equation of motion for $\xi$ is then $L_z - \ell_z =0$. 

Since the system is gauge invariant we can fix the gauge by imposing a
condition of the form $\G(\{\mat{r}_\alpha\},\xi) = 0$, which is
equivalent to selecting a rotating frame in which the trajectory of
the system $(\{\mat{r}_\alpha(t)\},\xi(t))$ in configuration space is
constrained to satisfy the relation $\G(\{\mat{r}_\alpha(t)\},\xi(t))
= 0$.  The function $\G$ can be chosen arbitrarily, as long as any
trajectory $(\{\mat{r}_\alpha^\prime\},\xi^\prime)$ can be transformed
into a new one $(\{\mat{r}_\alpha\},\xi)$ satisfying $\G=0$.  The new
trajectory must be unique, in the sense that no other trajectory
obtained from $(\{\mat{r}_\alpha^\prime\},\xi^\prime)$ by a gauge
transformation satisfies the gauge condition. Otherwise the gauge is
said to be ambiguous \cite{grb}.  Supplementary conditions must then
be imposed to fix the ambiguity.

\subsection{The laboratory frame}
\label{sec:labo}

Given any trajectory of the system $(\{\mat{r}_\alpha(t)\},\xi(t))$ by
means of a finite gauge transformation $\mat{r}_\alpha^\prime =
\mat{U}(\theta(t)) \mat{r}_\alpha$, $\xi^\prime(t) = \xi(t) -
\dot{\theta}(t)$, with \mat{U}\ an orthogonal $2\times2$ matrix and
$\theta(t) = \int_{t_0}^td\tau \xi(\tau)$, we can obtain a physically
equivalent trajectory with $\xi^\prime(t) = 0$.  The gauge
condition $\xi = 0$ corresponds to choosing a non-rotating reference
frame, the laboratory frame.  In this gauge the Lagrangian
(\ref{eq:glag}) reduces to (\ref{eq:rlag}).  The equation of motion
for $\xi$, $\sum_{\alpha=1}^N m_\alpha \mat{r}_\alpha \wedge
\dot{\mat{r}}_\alpha - \ell_z =0$, which cannot be obtained from
(\ref{eq:rlag}), must be imposed on the system as a constraint
\cite{lee}.  In the Hamiltonian formulation in this gauge that
constraint is first-class \cite{dir}, not leading to further
secondary constraints.

The quantization in the gauge $\xi=0$ is canonical.  In units such
that $\hbar=1$ we have,
\begin{equation}
  \label{eq:xi0}
    \H = \sum_{\alpha=1}^N \frac{1}{2 m_\alpha} \mat{p}_\alpha^2 + \V,
\quad
    \left[ r_{\alpha i}, p_{\beta j} \right]  = i \delta_{\alpha\beta}
    \delta_{ij},  
\quad
      \langle \phi | \psi \rangle  = \int \prod_{\beta=1}^N
      d^2\mat{r}_\beta
      \phi^*\left(\left\{\mat{r}_\alpha\right\}\right)
    \psi\left(\left\{\mat{r}_\alpha\right\}\right),
\end{equation}
with $\mat{p}_\alpha = 1/i \mat{\nabla}_{\alpha}$. The first-class
constraint is imposed on the state space \cite{dir}, $L_z |\psi\rangle
= \ell_z |\psi\rangle$.  The constant $\ell_z$ can only take integer
values in the quantum theory.  We see that both the classical and
quantum theories for this model in the gauge $\xi=0$ are completely
analogous to electrodynamics in Weyl gauge \cite{lee,kiw} (see
appendix \ref{sec:appqed}).  The constraint fixing the value of $L_z$,
in particular, plays the same role as Gauss law in QED.

\section{Linear gauge conditions}
\label{sec:linear}

The simplest form of gauge condition involving the coordinates is a
linear relation among them.  As discussed in section \ref{sec:eckart}
this kind of gauge condition is relevant in the context of
perturbative or semiclassical expansions.  The following notations
will be used throughout this paper,
\begin{subequations}
\label{eq:lng+}
  \begin{gather}
  \label{eq:lng}
  \S(\{\mat{r}_\alpha\}) = \sum_{\beta =1}^N m_\beta
  \left( A_\beta x_\beta + B_\beta y_\beta \right),\\
  \Q(\{\mat{r}_\alpha\}) = \sum_{\alpha =1}^N m_\alpha
  \left( B_\alpha x_\alpha - A_\alpha y_\alpha \right),
  \quad
  \R^2 \equiv \sum_{\alpha =1}^N m_\alpha \left( A_\alpha^2 +
  B_\alpha^2 \right).
  \end{gather}
\end{subequations}
The general linear gauge condition is of the form $\S = 0$, with $\R^2
\neq 0$.  We denote position vectors $\mat{R}_\alpha$ and their
components $X_\alpha$, $Y_\alpha$ in this gauge by capital letters, as
opposed to vectors in the gauge $\xi =0$ (the laboratory frame)
denoted by $\mat{r}_\alpha$.  Thus $\S(\{\mat{R}_\alpha\}) = 0$ but,
in general, $\S(\{\mat{r}_\alpha\}) \neq 0$.  This gauge condition
selects a reference frame rotating in such a way that the linear
combination of coordinates \S\ vanishes for all $t$.  If we choose,
for instance, all coefficients in (\ref{eq:lng}) vanishing except for
$B_1$, the coordinate frame must rotate together with particle 1 so
that particle stays on the $X$ axis for all $t$.  The formalism in
these linear gauges is entirely analogous to that of electrodynamics
in Coulomb gauge (appendix \ref{sec:appqed}), in which the fields are
also constrained by a linear relation.

The transformation from the gauge $\xi=0$ to the gauge $\S =0$ is
given by,
\begin{equation}
  \label{eq:tht}
  \mat{R}_\alpha(t) = \mat{U}(\theta(t)) \mat{r}_\alpha (t),
  \quad
  \xi(t) = - \dot{\theta}(t),
  \quad \text{with} \quad 
  \theta (t) = \mathrm{arctan} \left(\frac{\s}{\q}\right) + n \pi,
\end{equation}
where we denoted $\s \equiv \S(\{\mat{r}_\alpha\})$, $\q \equiv
\Q(\{\mat{r}_\alpha\})$ for brevity.  The indetermination in $\theta$
up to addition of $\pi$ is a Gribov ambiguity \cite{grb} (see also
\cite{lee,chr}), related to the two possible choices $\Q = \pm
\sqrt{\q^2 + \s^2}$.  We fix the ambiguity by requiring $\Q \geq 0$
and $-\pi < \theta \leq \pi$.  Due to the relation $\xi =
-\dot{\theta}$ for all $t$ in this gauge, with $\theta$ from
(\ref{eq:tht}), we can use $\{\mat{R}_\alpha, \theta\}$ as dynamical
variables instead of $\{\mat{R}_\alpha, \xi\}$.  The former set of
variables is preferable in the operator approach we espouse in this
paper, whereas in the path integral formulation switching from one set
to the other amounts to a mere change of integration variables.  The
inverse to the transformation (\ref{eq:tht}) is then,
\begin{equation}
  \label{eq:ith}
  \mat{r}_\alpha(t) = \mat{U}(-\theta(t)) \mat{R}_\alpha (t),
\end{equation}
with $\theta$ an independent variable. (If we choose $\xi$ as a
dynamical variable instead of $\theta$, then in (\ref{eq:ith})
$\theta(t) = - \int_{t_0}^t d\tau \xi(\tau)$.)

Solving the equation of motion for $\xi$ we get,
\begin{equation}
  \label{eq:xi}
  \xi = - \dot{\theta} = \frac{1}{\sum_{\beta=1}^N m_\beta
  \mat{R}_\beta^2} \left(\ver{z} \cdot \sum_{\alpha=1}^N
  m_\alpha \mat{R}_\alpha \wedge \dot{\mat{R}}_\alpha - \ell_z\right). 
\end{equation}
Since the gauge has already been completely fixed, we can substitute
(\ref{eq:xi}) back into the Lagrangian (\ref{eq:glag}).  Notice that
the vector product appearing in (\ref{eq:xi}) is not the total angular
momentum of the system.  In the classical theory we obtain the momenta
$\mat{\varPi}_\alpha$ conjugate to $\mat{R}_\alpha$ by differentiating
\Ll\ in (\ref{eq:glag}) with respect to $\dot{\mat{R}}_\alpha$ under
the constraint $\dot{\S}(\{\mat{R}_\alpha \}) =
\S(\{\dot{\mat{R}}_\alpha\}) = 0$, to obtain,
\begin{equation}
  \label{eq:mom}
  \varPi_{X_\alpha} = m_\alpha \dot{X}_\alpha + m_\alpha \xi \left( Y_\alpha
  + \frac{A_\alpha \Q}{\R^2} \right),
\qquad
  \varPi_{Y_\alpha} = m_\alpha \dot{Y}_\alpha - m_\alpha \xi \left( X_\alpha
  - \frac{B_\alpha \Q}{\R^2} \right),
\end{equation}
where $\xi$ is given by (\ref{eq:xi}). These momenta are consistent
with the gauge condition, since they satisfy 
\begin{equation}\label{eq:rel}
  0 = \sum_{\beta=1}^N \frac{1}{m_\beta} \frac{\partial \S}{\partial
  \mat{R}_\beta} \cdot \mat{\varPi}_\beta =
  \S\left(\left\{\mat{\varPi}_\alpha/m_\alpha\right\}\right), 
\end{equation}
the last equality following from the linearity of $\S$.  Relation
(\ref{eq:rel}) is analogous to the condition that the momentum
conjugate to the potential in Coulomb gauge must be transverse, eq.\
(\ref{eq:em9+}). 

From (\ref{eq:tht}) and (\ref{eq:ith}) we can obtain the relation
between the velocities $\{\dot{\mat{r}}_\alpha\}$ in the gauge $\xi=0$
and those in the gauge $\S=0$, $\{\dot{\mat{R}}_\alpha, \dot{\theta}\}$.
Correspondingly, we can express the momenta $\{\mat{p}_\alpha\}$ in
one gauge in terms of momenta $\{\mat{\varPi}_\alpha\}$ and $\ell_z$
in the other.  With those transformations we obtain from \H\ in
(\ref{eq:xi0}) the classical Hamiltonian in this gauge,
\begin{equation}
  \label{eq:ham}
  \H = \sum_{\alpha=1}^N \frac{1}{2m_\alpha} \mat{\varPi}_\alpha^2
  + \frac{\R^2}{2\Q^2} (\ell_z - \Lambda)^2 + \V,
  \quad \text{with} \quad \Lambda = \sum_{\beta=1}^N \left(
  X_\beta \varPi_{Y_\beta} - Y_\beta \varPi_{X_\beta} \right).
\end{equation}
The quantity $\Lambda$ defined by this equation will be henceforth
referred to as the ``residual angular momentum.''

By the same token, expressing $\mat{\varPi}_\alpha$ and
$\mat{R}_\alpha$ in terms of $\mat{p}_\alpha$ and $\mat{r}_\alpha$ and
using the Poisson brackets (\ref{eq:xi0}) we get the Poisson brackets
in this gauge.  Alternatively, they can be found as Dirac brackets
\cite{dir} relative to the second-class constraints
$\S(\{\mat{R}_\alpha\}) = 0$ $=\S(\{\mat{\varPi}_\alpha/m_\alpha\})$.
The result is, written in the notation of quantum commutators,
\begin{equation}
  \label{eq:cmm}
\begin{gathered}
  \left[X_\beta,\varPi_{X_\gamma}\right] = i\left(\delta_{\beta\gamma}
    - \frac{A_\beta A_\gamma m_\gamma}{\R^2}\right), \quad
  \left[Y_\beta,\varPi_{Y_\gamma}\right] = i\left(\delta_{\beta\gamma}
    - \frac{B_\beta B_\gamma m_\gamma}{\R^2}\right),
  \\
  \left[X_\beta,\varPi_{Y_\gamma}\right] = -i
  \frac{A_\beta B_\gamma m_\gamma}{\R^2}, \quad
  \left[Y_\beta,\varPi_{X_\gamma}\right] = -i
  \frac{m_\gamma A_\gamma B_\beta}{\R^2}.
\end{gathered}
\end{equation}
All other commutators among components of $\mat{R}_\alpha$ and
$\mat{\varPi}_\beta$ vanish.  The correspondence between
(\ref{eq:cmm}) and (\ref{eq:em12}) is apparent.  Using (\ref{eq:cmm})
we obtain the commutators for $\Lambda$,
\begin{equation}
  \label{eq:+cmm}
  \begin{gathered}
    \left[X_\alpha, \Lambda\right] = -i Y_\alpha - i \frac{A_\alpha
      \Q}{\R^2}, \qquad
    \left[Y_\alpha, \Lambda\right] = i X_\alpha - i \frac{B_\alpha
      \Q}{\R^2},\\ 
    \left[\Lambda,\varPi_{X_\gamma}\right] = i \varPi_{Y_\gamma} + i
    \frac{A_\gamma m_\gamma}{\R^2}
    \Q\left(\{\mat{\varPi}_\alpha/m_\alpha\}\right), \quad 
    \left[\Lambda,\varPi_{Y_\gamma}\right] = -i \varPi_{X_\gamma} + i
    \frac{B_\gamma m_\gamma}{\R^2}
    \Q\left(\{\mat{\varPi}_\alpha/m_\alpha\}\right). 
  \end{gathered}
\end{equation}
Furthermore, from (\ref{eq:cmm}) and (\ref{eq:+cmm}) we get,
\begin{equation}
  \label{eq:++cmm}
  \begin{gathered}[]
  [\S, \varPi_{X_\alpha}] = [\S, \varPi_{Y_\alpha}] = 0 = [\S,
  \Lambda] = [\S\left(\left\{ \mat{\varPi}_\alpha/m_\alpha
  \right\}\right), \Lambda], \\ 
  [\Q,\varPi_{X_\alpha}] = i m_\alpha B_\alpha,
  \quad
  [\Q,\varPi_{Y_\alpha}] = -i m_\alpha A_\alpha,
  \quad
  [\Q, \S\left(\left\{ \mat{\varPi}_\alpha/m_\alpha
    \right\}\right)] = 0,
  \quad
  [\Q, \Lambda] = -i \S.
  \end{gathered}
\end{equation}
We see from (\ref{eq:cmm})--(\ref{eq:++cmm}) that
$\mat{\varPi}_\alpha$ and $\Lambda$ generate translations and
rotations, respectively, of $\{\mat{R}_\alpha\}$ on the surface $\S =
0$.  We see also that the gauge condition $\S =0$ and (\ref{eq:rel})
are operator equations, that can be evaluated within commutators.

In the quantum theory a realization of the commutators (\ref{eq:cmm})
is obtained by defining $\mat{\varPi}_\alpha$ as the projection of the
gradient $\mat{\nabla}_\alpha$ on the hyperplane tangent to the
surface $\S = 0$ (which in this case is the surface itself, since \S\ 
is a linear function),
\begin{equation}
  \label{eq:qmom}
  \begin{split}
    \varPi_{X_\alpha} &= \frac{1}{i} \frac{\partial}{\partial
    X_\alpha} - \frac{1}{i} \frac{m_\alpha A_\alpha}{\R^2}
    \sum_{\beta=1}^N \left( A_\beta \frac{\partial}{\partial X_\beta}
      + B_\beta \frac{\partial}{\partial Y_\beta} \right),\\
    \varPi_{Y_\alpha} &= \frac{1}{i} \frac{\partial}{\partial
    Y_\alpha} - \frac{1}{i} \frac{m_\alpha B_\alpha}{\R^2}
    \sum_{\beta=1}^N \left( A_\beta \frac{\partial}{\partial X_\beta}
      + B_\beta \frac{\partial}{\partial Y_\beta} \right).
  \end{split}
\end{equation}
These operators, which are analogous to (\ref{eq:em14}) in QED in
Coulomb gauge, satisfy the equation
$\S(\{\mat{\varPi}_\alpha/m_\alpha\}) = 0$.  From (\ref{eq:ham}) and
(\ref{eq:qmom}) we obtain the expression for the residual angular
momentum operator $\Lambda$,
\begin{equation}
  \label{eq:intri}
  \Lambda = \sum_{\beta=1}^N \left\{ \left( X_\beta - \frac{B_\beta
        \Q}{\R^2} \right) \frac{1}{i} \frac{\partial}{\partial
        Y_\beta} - \left( Y_\beta + \frac{A_\beta
        \Q}{\R^2} \right) \frac{1}{i} \frac{\partial}{\partial
        X_\beta} \right\},
\end{equation}
which is shown in appendix \ref{sec:appL} to have integer eigenvalues.

Using the relations (\ref{eq:tht}) and (\ref{eq:ith}) between
$\{\mat{r}_\alpha\}$ and $\{\mat{R}_\alpha,\theta\}$ and applying the
chain rule we obtain, after appropriately rearranging the derivative
operators, 
\begin{equation}\label{eq:zine}
\begin{split}
  p_{x\alpha} & \equiv \frac{1}{i} \frac{\partial}{\partial x_\alpha}
  = \cos\theta \left(\varPi_{X\alpha} + \frac{m_\alpha A_\alpha}{\Q}
    (L_z - \Lambda)\right) - \sin\theta \left(\varPi_{Y\alpha} +
    \frac{m_\alpha B_\alpha}{\Q} (L_z - \Lambda)\right), \\
  p_{y\alpha} & \equiv \frac{1}{i} \frac{\partial}{\partial y_\alpha}
  = \sin\theta \left(\varPi_{X\alpha} + \frac{m_\alpha A_\alpha}{\Q}
    (L_z - \Lambda)\right) + \cos\theta \left(\varPi_{Y\alpha} +
    \frac{m_\alpha B_\alpha}{\Q} (L_z - \Lambda)\right),
\end{split}
\end{equation}
with $L_z = 1/i\,\partial/\partial\theta$.  The first-class constraint
$L_z \psi = \ell_z \psi$
is trivial to solve in this gauge, $\psi(\{\mat{R}_\alpha\},\theta) =
\psi(\{\mat{R}_\alpha\}) \exp(i \ell_z\theta) / \sqrt{2\pi}$.  An
analog of (\ref{eq:zine}) in QED is the simpler relation
(\ref{eq:em17}). 

The Hamiltonian operator in this gauge is obtained from \H\ in the
gauge $\xi=0$ as given by (\ref{eq:xi0}), through the transformation
rules (\ref{eq:ith}) and (\ref{eq:zine}),
\begin{equation}
  \label{eq:qham}
  \H = \sum_{\beta=1}^N \frac{1}{2 m_\beta} \left( \frac{1}{\Q}
  \varPi_{X_\beta} \Q \varPi_{X_\beta} + \frac{1}{\Q} \varPi_{Y_\beta}
  \Q \varPi_{Y_\beta} \right) + \frac{\R^2}{2\Q^2} (\ell_z -
  \Lambda)^2 + \V.
\end{equation}
The first term in \H\ has a structure similar to that of the Laplacian
in curvilinear coordinates, with \Q\ as the Jacobian and \mat{\varPi}\ 
as derivative operators. (In the case $N=1$ the similarity turns, in
fact, into an identity, see below.) Notice, however, that we did not
postulate (\ref{eq:qham}), rather, we derived it from the expression
(\ref{eq:xi0}) in the lab frame.

The inner product in Hilbert space can be found from the expression
(\ref{eq:xi0}) for $\langle \phi | \psi \rangle$ in the gauge $\xi =
0$ by the familiar Faddeev-Popov method \cite{fad}, the relevant
resolution of the identity being in this case,
\begin{equation}
  \label{eq:fad}
  1 = \int_{-\pi}^\pi d\alpha\; \delta\left(
  \S\left( \{\mat{U}(\alpha) \mat{r}_\beta \}\right)\rule{0pt}{2ex}
  \right) \Theta\left(\Q\left( \{\mat{U}(\alpha) \mat{r}_\beta
  \}\right) \rule{0pt}{2ex} \right) \Q(\{\mat{U}(\alpha) \mat{r}_\beta\}), 
\end{equation}
with $\Theta(\Q)$ a step function enforcing positivity of the
Faddeev-Popov determinant $\Q$. As mentioned above, the condition $\Q
\geq 0$ guarantees that there is only one root $\alpha = \theta$ (with
$\theta$ from (\ref{eq:tht})), and not $\theta + \pi$, to the equation
$\S=0$ in (\ref{eq:fad}).  Inserting (\ref{eq:fad}) in the expression
(\ref{eq:xi0}) for $\langle \phi | \psi \rangle$ we get,
\begin{equation}
  \label{eq:inp}
  \langle \phi | \psi \rangle = \int \prod_{\beta =1}^N
  d^2\mat{R}_\beta \delta(\S) \Theta(\Q)
  \Q \phi^* (\{\mat{R}_\alpha\}) \psi(\{\mat{R}_\alpha\}), 
\end{equation}
where we dropped a factor of $2\pi$, the measure of the group $SO(2)$.
The hermitianity of $\Lambda$ and $\V$ with respect to the inner
product (\ref{eq:inp}) is immediate in view of the commutation
relations.  In order to check the hermitianity of the Hamiltonian
(\ref{eq:qham}) it is then enough to show that the first term in
(\ref{eq:qham}) is hermitian.  That calculation is straightforward,
though somewhat lengthy, so we omit the details.

It is sometimes convenient to redefine the state space by absorbing
the Jacobian in the wave functions and eliminating it from the
integration measure in the inner product.  The redefined wave
functions are $\widetilde{\psi} = \Q^{1/2} \psi$, leading to the
Hamiltonian,
\begin{equation}
  \label{eq:nham}
  \widetilde{\H} \equiv \Q^{1/2} \H \Q^{-1/2} = \sum_{\beta=1}^N
  \frac{1}{2m_\beta} \left(\varPi_{X_\beta}^2 + \varPi_{Y_\beta}^2 
  \right) + \frac{\R^2}{2\Q^2} (\ell_z - \Lambda)^2 + \V -
  \frac{\R^2}{8\Q^2},
\end{equation}
the last term being the quantum-mechanical potential \cite{lee,chr}.
Equation (\ref{eq:nham}), with classical \mat{\varPi}\ and $\Lambda$,
is the Hamiltonian found in this gauge in the path-integral approach.
Since the transformation (\ref{eq:tht}) depends non-linearly on
$\{\mat{r}_\alpha\}$, the associated change of integration variables
in the generating functional entails a change in its discretization
\cite{tir}, which ultimately gives rise to the quantum potential term.

\subsection{The case \mat{N=1}}
\label{sec:n=1}

The case $N=1$ is instructive \cite{lee,chr}.  By means of a
time-independent rotation we can always reduce the condition $\S = 0$
to $Y =0$.  We thus fix a reference frame rotating together with the
particle, so that it is on the $X$ axis for all $t$, with $X \geq 0$.
In order to specify the position of the particle, we give its
coordinate $X$ and the angle $\theta$ of the $X$ axis relative to the
laboratory $x$ axis.  We are then describing the motion in terms of
polar coordinates with $X$ the radial coordinate.

From (\ref{eq:qmom}) we have $\varPi_X = -i \partial/\partial X$ and
$\varPi_Y = 0 = \Lambda$.  The Faddeev-Popov determinant in this case
is $\Q =X$, and the Hamiltonian (\ref{eq:qham}) reduces to that of a
particle in polar coordinates, with angular momentum $\ell_z$.
Similarly, the inner product (\ref{eq:inp}) corresponds to polar
coordinates.  Due to the constraint $L_z \psi =\ell_z \psi$ the
integration over the angle variable $\theta$ is trivial, so only the
radial wave function appears in (\ref{eq:inp}).  If the wave function
in Cartesian coordinates is $\psi(x,y)$, the radial wave function is
$\psi(X,0)$. The quantum potential term in (\ref{eq:nham}) also
reduces in this case to its well-known form \cite{lee,chr} for polar
coordinates, $-1/(8mX^2)$.

\section{Quadratic gauge condition: the instantaneous principal axes}  
\label{sec:quadratic}

The quantization of the system (\ref{eq:rlag}) in a rotating frame
defined by a quadratic gauge condition follows the same lines as the
linear case studied in section \ref{sec:linear}.  Both the treatment
and the results are technically more involved, however, so instead of
considering a general quadratic gauge condition we restrict ourselves
to the particular case of the instantaneous principal axes frame.
That reference frame plays a central role in the treatment of rigid
body dynamics.  In the case of many-body systems, their quantization
in the instantaneous principal axes frame has been proposed as a
method for separating the ``collective'' rotations from the
``intrinsic'' dynamics.  We briefly discuss that issue in section
\ref{sec:final}.  In this section we compute the quantum Hamiltonian
and inner product by means of a gauge transformation from the gauge
$\xi = 0$.

We define the quantities,
\begin{equation}
  \label{eq:qquan}
  Q(\{\mat{r}_\alpha\}) = \frac{1}{2} \sum_{\alpha=1}^N m_\alpha
  (x_\alpha^2 - y_\alpha^2),
  \quad
  S(\{\mat{r}_\alpha\}) = \sum_{\alpha=1}^N m_\alpha x_\alpha
  y_\alpha,
  \quad
  R^2(\{\mat{r}_\alpha\}) = \sum_{\alpha=1}^N m_\alpha
  \mat{r}_\alpha^2. 
\end{equation}
$R^2$ is the trace of the inertia tensor of the system, whose
traceless part is given by $\left( \begin{smallmatrix} -Q & -S\\-S & Q
  \end{smallmatrix} \right)$. 
The instantaneous principal axes frame
is then defined by the condition $S=0$.  As above, we denote vectors
referred to this frame with capital letters, so that
$S(\{\mat{R}_\alpha\})=0$.  The gauge transformation from the gauge
$\xi =0$ to the gauge $S=0$ has the form (\ref{eq:tht}), with the
parameter,
\begin{equation}
  \label{eq:qtht}
  \theta(t) = \frac{1}{2} \mathrm{arctan}\left(
  \frac{S(\{\mat{r}_\alpha\})}{Q(\{\mat{r}_\alpha\})}  \right) +
  \frac{n}{2} \pi.
\end{equation}
Due to the fact that the inertia tensor is second rank the number of
Gribov ambiguities doubles with respect to the linear case
(\ref{eq:tht}), there being now four solutions in the range $-\pi \leq
\theta \leq \pi$.  We fix the ambiguity by requiring
$Q(\{\mat{R}_\alpha\}) \geq 0$ and $0 \leq \theta \leq \pi$.

The Lagrangian is now (\ref{eq:glag}) with $\xi$ having the same form
as in (\ref{eq:xi}).  Consistency with the gauge condition
$S(\{\mat{R}_\alpha\}) = 0$ requires the velocities and momenta to
satisfy, 
  \begin{subequations}
\begin{align}
  \label{eq:ds}
  \sum_{\alpha
    =1}^N m_\alpha (X_\alpha \dot{Y}_\alpha + Y_\alpha \dot{X}_\alpha)
  & = 0,
\\
  \label{eq:ds*}
  \sum_{\alpha =1}^N (X_\alpha \varPi_{Y_\alpha} + Y_\alpha
  \varPi_{X_\alpha}) & = 0.  
\end{align}
  \end{subequations}
We obtain the conjugate momenta in terms of velocities by deriving \Ll\
with respect to $\dot{\mat{R}}_\alpha$ under the constraint
(\ref{eq:ds}),
\begin{equation}
  \label{eq:cmom}
  \varPi_{X_\alpha} = m_\alpha \dot{X}_\alpha + \xi m_\alpha Y_\alpha
    \left( 1 + \frac{2 Q(\{\mat{R}_\alpha\})}{R^2(\{\mat{R}_\alpha\})} 
  \right),
  \quad
  \varPi_{Y_\alpha} = m_\alpha \dot{Y}_\alpha - \xi m_\alpha X_\alpha
    \left( 1 - \frac{2 Q(\{\mat{R}_\alpha\})}{R^2(\{\mat{R}_\alpha\})} 
  \right).  
\end{equation}
The basic Poisson brackets in this gauge can be obtained as in section
\ref{sec:linear}.  Since the momenta $\mat{\varPi}_\alpha$ generate
translations on the curved hypersurface $S(\{\mat{R}_\alpha\}) = 0$,
which do not commute, the Poisson brackets among momenta do not
vanish.  Correspondingly, in the quantum theory the operators
$\mat{\varPi}_\alpha$ do not commute with each other.  The
non-vanishing commutators among coordinates and momenta are,
\begin{gather}
  \left[X_\beta,\varPi_{X_\gamma}\right] = i\left(\delta_{\beta\gamma}
    - \frac{Y_\beta Y_\gamma m_\gamma}{R^2}\right), 
  \quad
  \left[Y_\beta,\varPi_{Y_\gamma}\right] = i\left(\delta_{\beta\gamma}
    - \frac{X_\beta X_\gamma m_\gamma}{R^2}\right), 
  \nonumber\\
  \left[X_\beta,\varPi_{Y_\gamma}\right] = -i \frac{Y_\beta X_\gamma
    m_\gamma}{R^2},  \label{eq:qcomm}
  \quad
  \left[Y_\beta,\varPi_{X_\gamma}\right] = -i
  \frac{X_\beta Y_\gamma m_\gamma }{R^2},
  \\
  \left[\varPi_{X_\beta},\varPi_{X_\gamma}\right] = \frac{i}{R^2}
    \left( m_\gamma Y_\gamma \varPi_{Y_\beta} - m_\beta Y_\beta 
      \varPi_{Y_\gamma} \right),
  \nonumber\\
  \left[\varPi_{X_\beta},\varPi_{Y_\gamma}\right] = \frac{i}{R^2}
    \left( m_\gamma X_\gamma \varPi_{Y_\beta} - m_\beta Y_\beta 
      \varPi_{X_\gamma} \right),
\quad
  \left[\varPi_{Y_\beta},\varPi_{Y_\gamma}\right] = \frac{i}{R^2}
    \left( m_\gamma X_\gamma \varPi_{X_\beta} - m_\beta X_\beta 
      \varPi_{X_\gamma} \right).\nonumber
\end{gather}
Like in the previous section, we obtain a realization of this
commutator algebra in terms of differential operators by projecting
the gradient operator on the hyperplane tangent to $S=0$,
\begin{equation}
  \label{eq:qmom2}
  \begin{split}
    \varPi_{X_\alpha} &= \frac{1}{i} \frac{\partial}{\partial
    X_\alpha} - \frac{1}{i} \frac{m_\alpha Y_\alpha}{R^2}
    \sum_{\beta=1}^N \left( Y_\beta \frac{\partial}{\partial X_\beta}
      + X_\beta \frac{\partial}{\partial Y_\beta} \right),\\
    \varPi_{Y_\alpha} &= \frac{1}{i} \frac{\partial}{\partial
    Y_\alpha} - \frac{1}{i} \frac{m_\alpha X_\alpha}{R^2}
    \sum_{\beta=1}^N \left( Y_\beta \frac{\partial}{\partial X_\beta}
      + X_\beta \frac{\partial}{\partial Y_\beta} \right).
  \end{split}
\end{equation}
Both the classical momenta (\ref{eq:cmom}) and the quantum operators
(\ref{eq:qmom2}) satisfy the constraint (\ref{eq:ds*}).  The gauge
condition $S(\{\mat{R}_\alpha\}) = 0$ and its counterpart
(\ref{eq:ds*}) are operator equations, that can be evaluated inside
commutators as can be easily checked from (\ref{eq:qcomm}).  The
Hamiltonian operator is written in terms of momentum operators
(\ref{eq:qmom2}) as,
\begin{equation}
  \label{eq:cham}
  \H = \sum_{\beta=1}^N \frac{1}{2m_\beta} \left( \frac{R}{2Q}
  \varPi_{X_\beta} \frac{2Q}{R} \varPi_{X_\beta} + \frac{R}{2Q}
  \varPi_{Y_\beta} \frac{2Q}{R} \varPi_{Y_\beta} \right) +
  \frac{R^2}{8Q^2} (\ell_z -\Lambda)^2 + \V,
\end{equation}
with $R = \sqrt{R^2}$ and 
\begin{equation}
  \label{eq:cint}
  \Lambda \equiv \sum_{\alpha =1}^N \left( X_\alpha \varPi_{Y_\alpha} -
  Y_\alpha \varPi_{X_\alpha} \right) = \left( 1 - \frac{2 Q}{R^2}
  \right) \sum_{\alpha=1}^N X_\alpha \frac{1}{i}
  \frac{\partial}{\partial Y_\alpha} - \left( 1 + \frac{2
  Q}{R^2}\right) \sum_{\alpha=1}^N Y_\alpha \frac{1}{i}
  \frac{\partial}{\partial X_\alpha}
\end{equation}
the residual angular momentum in this gauge.  From (\ref{eq:cham}) we
find the form of the quantum potential, 
\begin{equation}
  \label{eq:cqpot}
  \V_Q = -\frac{R^2}{8 Q^2} + \frac{7-4 N}{8 R^2}.
\end{equation}
In order to find the inner product in this gauge we start from the
resolution of the identity,
\begin{equation}
  \label{eq:cfad}
  1 = \int_{0}^\pi d\alpha\; \delta\left(
  S\left( \{\mat{U}(\alpha) \mat{r}_\beta \}\right)\rule{0pt}{2ex}
  \right) \Theta\left(Q\left( \{\mat{U}(\alpha) \mat{r}_\beta
  \}\right) \rule{0pt}{2ex} \right) 2 Q(\{\mat{U}(\alpha)
  \mat{r}_\beta\}).  
\end{equation}
Notice that we restricted the integration range to $0\leq \alpha \leq
\pi$.  Alternatively, we can integrate from $-\pi$ to $\pi$, and set
the l.h.s.\ of (\ref{eq:cfad}) equal to 2.  The inner product is then
obtained as,
\begin{equation}
  \label{eq:cinp}
  \begin{aligned}
  \langle \phi | \psi \rangle &= \frac{1}{2}\int \prod_{\beta =1}^N 
  d^2\mat{R}_\beta \delta(S) \Theta(Q)
  2Q \phi^* (\{\mat{R}_\alpha\}) \psi(\{\mat{R}_\alpha\})\\ 
  & = \frac{1}{2}\int \prod_{\beta =1}^N 
  d^2\mat{R}_\beta \delta\left(\frac{S}{R}\right) \Theta(Q)
  \frac{2Q}{R} \phi^* (\{\mat{R}_\alpha\}) \psi(\{\mat{R}_\alpha\}), 
  \end{aligned}
\end{equation}
where we divided by $2\pi$, and omitted the argument
$\{\mat{R}_\alpha\}$ in $S$, $Q$ and $R$ for simplicity.  In the
second line we wrote the gauge condition as $S/R$, taking into account
that the factor $1/R$ has no finite zeros and that its singularity at
the origin is suppresed by the zeros of $S$ and $Q$ there.  This last
expression for $\langle \phi | \psi \rangle$ is convenient to check
the hermitianity of \H.  We again omit the details of that proof. 

As a consequence of the gauge condition $S(\{\mat{R}_\alpha\})=0$
being quadratic, the momentum operators $\mat{\varPi}_\alpha$ in
(\ref{eq:qmom2}) have coefficients which are ratios of quadratic
polynomials in the coordinates, whereas the operators (\ref{eq:qmom})
for a linear gauge have constant coefficients.  Accordingly, the basic
commutators (\ref{eq:qcomm}) are rational functions of coordinates
and, furthermore, momentum operators do not commute with each other.
The structure of the Hamiltonian operator in this gauge, eq.\ 
(\ref{eq:cham}), is also much more complicated than in linear gauges,
eq.\ (\ref{eq:qham}).  As shown in appendix \ref{sec:appL}, the
residual angular momentum operator $\Lambda$, (\ref{eq:cint}), does
not have integer eigenvalues.

\section{Center of mass motion}
\label{sec:cmm}

In this section and the next one we set $U = 0$ in the Lagrangian and
take into account the translation invariance of (\ref{eq:rlag}) in
order to separate the center of mass degrees of freedom.  Since the
motion of the center of mass is dynamically trivial, we restrict our
treatment to dynamical states with vanishing total momentum.  We
consider linear gauge conditions only.

The Lagrangian (\ref{eq:rlag}) is invariant under time-independent
transformations of the Euclidean group,
\begin{equation}
  \label{eq:eutr}
  \mat{r}^\prime_\alpha = \mat{U}(\theta) \mat{r}_\alpha + \mat{u}, 
\end{equation}
with \mat{U}\ an orthogonal matrix.  We define the covariant
derivative $D_t \mat{r}_\alpha = \dot{\mat{r}}_\alpha - \xi \ver{z}
\wedge \mat{r}_\alpha - \mat{\rho}$.  Under time-dependent
transformations $\mat{r}_\alpha$ transforms as in (\ref{eq:eutr})
and,
\begin{equation}
  \label{eq:eutr1}
  \xi^\prime = \xi - \dot{\theta},
  \quad
  \mat{\rho}^\prime = \mat{U}(\theta) \mat{\rho} + \dot{\mat{u}} - 
   ( \xi - \dot{\theta} ) \ver{z} \wedge \mat{u},
  \quad
  (D_t \mat{r}_\alpha)^\prime = \mat{U}(\theta) D_t \mat{r}_\alpha.
\end{equation}
Substituting $\dot{\mat{r}}_\alpha$ by $D_t \mat{r}_\alpha$ in
(\ref{eq:rlag}) we obtain a Lagrangian that is quasi-invariant under
the transformations (\ref{eq:eutr}), (\ref{eq:eutr1}), and which has
the form $\Ll + \Ll_\rho$, with \Ll\ given by (\ref{eq:glag}) and,
\begin{equation}
  \label{eq:rhola}
  \Ll_\rho = \frac{1}{2} \sum_{\alpha=1}^N m_\alpha \mat{\rho}^2 -
  \mat{\rho} \cdot \sum_{\alpha=1}^N m_\alpha \dot{\mat{r}}_\alpha +
  \xi \ver{z} \cdot \sum_{\alpha=1}^N m_\alpha \mat{r}_\alpha \wedge
  \mat{\rho}. 
\end{equation}
If we choose the gauge conditions $\xi = 0 = \mat{\rho}$ we recover
the Lagrangian (\ref{eq:rlag}), constrained by the eqs.\ of motion for
$\xi$ and \mat{\rho}\ in this gauge,
\begin{equation}
  \label{eq:oluc}
  \sum_{\alpha=1}^N m_\alpha \mat{r}_\alpha \wedge
  \dot{\mat{r}}_\alpha = \ell_z,
  \quad
  \sum_{\alpha=1}^N m_\alpha \dot{\mat{r}}_\alpha =0.
\end{equation}
These constraints are first class. In the quantum theory they restrict
the state space of the theory, $L_z \psi = \ell_z \psi$,
$\sum_{\alpha=1}^N \mat{\nabla}_\alpha \psi = 0$, analogously to Gauss
law (\ref{eq:em6b}) in QED.

We can now proceed along the same lines as in section
\ref{sec:linear}, imposing on the system the gauge conditions,
\begin{equation}
  \label{eq:gge}
  \S(\{\mat{R}_\alpha\}) = 0,
  \quad
  \mat{\C}(\{\mat{R}_\alpha\}) \equiv \frac{1}{M}\sum_{\beta=1}^N
  m_\beta \mat{R}_\beta = 0,
\end{equation}
with \S\ defined in (\ref{eq:lng}) and $M=\sum_{\alpha=1}^N m_\alpha$.
(\ref{eq:gge}) defines a reference frame in a particular state of
rotation, with origin at the center of mass.  Like in section
\ref{sec:linear} we denote vectors referred to this frame by capital
letters.  The gauge conditions (\ref{eq:gge}) are not mutually
consistent unless \S\ is translation invariant,
\begin{equation}
  \label{eq:trinv}
  \sum_{\alpha=1}^N m_\alpha A_\alpha = 0 = \sum_{\alpha=1}^N m_\alpha
  B_\alpha.  
\end{equation}
From the equations of motion for $\rho$ and $\xi$ in this gauge we
obtain $\mat{\rho}=0$ and the expression (\ref{eq:xi}) for $\xi$.  The
momenta conjugate to $\mat{R}_\alpha$ are given by (\ref{eq:mom}) in
terms of velocities. They satisfy (\ref{eq:rel}) and also
$\mat{\C}(\{\mat{\varPi}_\alpha/m_\alpha\}) = \sum_{\gamma=1}^N
\mat{\varPi}_\gamma =0$.  The classical Hamiltonian is given by
(\ref{eq:ham}).

The non-vanishing quantum commutators among coordinates and momenta in
this gauge are,
\begin{equation}
  \label{eq:cmm1}
\begin{gathered}
  \left[X_\beta,\varPi_{X_\gamma}\right] = i\left(\delta_{\beta\gamma}
    - \frac{A_\beta A_\gamma m_\gamma}{\R^2} -
    \frac{m_\gamma}{M}\right), 
  \quad
  \left[Y_\beta,\varPi_{Y_\gamma}\right] = i\left(\delta_{\beta\gamma}
    - \frac{B_\beta B_\gamma m_\gamma}{\R^2} - \frac{m_\gamma}{M}
  \right), 
  \\
  \left[X_\beta,\varPi_{Y_\gamma}\right] = -i
  \frac{A_\beta B_\gamma m_\gamma}{\R^2},
  \quad
  \left[Y_\beta,\varPi_{X_\gamma}\right] = -i
  \frac{m_\gamma A_\gamma B_\beta}{\R^2}.
\end{gathered}
\end{equation}
A realization of this algebra in terms of first order differential
operators with constant coefficients can be obtained as in
(\ref{eq:qmom}),
\begin{equation}
  \label{eq:qmom1}
  \begin{split}
    \varPi_{X_\alpha} &= \frac{1}{i} \frac{\partial}{\partial
      X_\alpha} - \frac{1}{i} \frac{m_\alpha}{M} \sum_{\beta=1}^N
    \frac{\partial}{\partial X_\beta} - \frac{1}{i} \frac{m_\alpha
      A_\alpha}{\R^2} \sum_{\beta=1}^N \left( A_\beta
      \frac{\partial}{\partial X_\beta}
      + B_\beta \frac{\partial}{\partial Y_\beta} \right),\\
    \varPi_{Y_\alpha} &= \frac{1}{i} \frac{\partial}{\partial
      Y_\alpha} - \frac{1}{i} \frac{m_\alpha}{M} \sum_{\beta=1}^N
    \frac{\partial}{\partial Y_\beta} - \frac{1}{i} \frac{m_\alpha
      B_\alpha}{\R^2} \sum_{\beta=1}^N \left( A_\beta
      \frac{\partial}{\partial X_\beta} + B_\beta
      \frac{\partial}{\partial Y_\beta} \right).
  \end{split}
\end{equation}
These operators satisfy the constraints
$\S(\{\mat{\varPi}_\alpha/m_\alpha\}) = 0 =
\mat{\C}(\{\mat{\varPi}_\alpha/m_\alpha\})$.  The realization of the
residual angular momentum as a differential operator is the same one
as in section \ref{sec:linear}, eq.\ (\ref{eq:intri}), since the extra
terms in (\ref{eq:qmom1}) with respect to (\ref{eq:qmom}) do not
contribute to $\Lambda = \sum_{\alpha=1}^N (X_\alpha \varPi_{Y_\alpha}
- Y_\alpha \varPi_{X_\alpha})$ due to the gauge condition $\mat{\C}
=0$. The Hamiltonian operator has the same form as in (\ref{eq:qham}),
but now with the momentum operators $\mat{\varPi}_\alpha$ from
(\ref{eq:qmom1}).  The inner product obtained with the Faddeev-Popov
procedure is,
\begin{equation}
  \label{eq:inp1}
  \langle \phi | \psi \rangle = \int \prod_{\beta =1}^N
  d^2\mat{R}_\beta \delta(\S) \delta^{(2)}(\mat{\C}) \Theta(\Q) 
  \Q \phi^*(\{\mat{R}_\alpha\}) \psi(\{\mat{R}_\alpha\}). 
\end{equation}
The quantum mechanical potential in this case is the same as in
(\ref{eq:nham}). 

\section{Quasirigid systems and the Eckart frame}
\label{sec:eckart}

We assume now that the potential energy \V\ (with $U =0$) has a
minimum for some configuration $\{\mat{z}_\alpha\}$ of the system,
such that $\mat{z}_\alpha \neq \mat{z}_\beta$ for some $\alpha \neq
\beta$, and that $\V_0 \equiv \V(\{\mat{z}_\alpha\}) \leq
\V(\{\mat{r}_\gamma\})$ for all configurations $\{\mat{r}_\gamma\}$.
Due to the invariance of \V\ under the Euclidean group $E_2$ any
configuration $\{\mat{z}_\alpha^\prime\}$ related to
$\{\mat{z}_\alpha\}$ by a transformation of the form (\ref{eq:eutr})
is also a minimum.  Denoting by $\mathcal{M}_\V$ the manifold of
configuration space defined by $\V(\{\mat{r}_\gamma\}) = \V_0$, we
assume that the quotient $\mathcal{M}_\V/E_2$ is a discrete set.  The
configurations of minimal potential energy are therefore rigid.  In
this section we discuss the quantization of the small oscillations of
the system about these rigid equilibrium configurations.  We will
denote by $\{\mat{Z}_\alpha\}$ the unique (up to discrete degeneracy)
minimum of \V\ satisfying,
\begin{equation}
  \label{eq:mini}
  \sum_{\alpha=1}^N m_\alpha Z_{\alpha x} Z_{\alpha y} = 0,
  \qquad
  \sum_{\alpha=1}^N m_\alpha \mat{Z}_{\alpha} = 0.
\end{equation}
The small oscillations of the system are therefore described by
trajectories of the form,
\begin{equation}
  \label{eq:traj}
  \mat{r}_\alpha (t) = \mat{z}_\alpha (t) + \mat{\delta r}_\alpha (t)
  \quad
  \text{with}
  \quad
  \mat{z}_\alpha (t) = \mat{U}(t) \mat{Z}_\alpha + \mat{u}
\end{equation}
for some orthogonal matrix $\mat{U}(t)$ and $\mat{u}$ appropriately
chosen so that $\mat{\delta r}_\alpha (t)$ are small with respect to
their characteristic scale for all $t$.  We do not assume, however,
that the velocities $\mat{\delta}\dot{\mat{r}}_\alpha$ are small.
Since we restrict ourselves to states with vanishing total momentum,
the translation vector $\mat{u}$ in (\ref{eq:traj}) must be
time-independent.

It is convenient to apply the inverse of the gauge transformation
defined by the second equation in (\ref{eq:traj}) in order to switch
to a reference frame, the ``body frame'' of the rigid equilibrium
configuration, so that
\begin{equation}
  \label{eq:body}
  \mat{r}_\alpha (t) = \mat{Z}_\alpha + \mat{\delta r}_\alpha (t).
\end{equation}
This fixes the gauge only to leading order in $\mat{\delta
r}_\alpha$.  We fix the residual gauge freedom by imposing a gauge
condition on $\mat{\delta r}_\alpha$, which amounts to correcting the
definition (\ref{eq:body}) of the reference frame by small quantities
of first order.  We choose the origin of the reference frame at the
center of mass, so to first order in $\mat{\delta r}_\alpha$ the gauge
conditions must be of the form (\ref{eq:gge}).  The choice of the
coefficients $A_\alpha$, $B_\alpha$ is arbitrary as long as
(\ref{eq:trinv}) is satisfied.  We then have,
\begin{equation}
  \label{eq:ppale}
  \mat{R}_\alpha (t) = \mat{Z}_\alpha + \mat{\delta R}_\alpha (t),
  \quad
  \S(\{\mat{\delta R}_\alpha\}) = 0,
  \quad
  \mat{\C}(\{\mat{\delta R}_\alpha\}) = 0.
\end{equation}
The instantaneous principal axes frame of section \ref{sec:quadratic},
for example, is defined to first order in $\mat{\delta R}_\alpha$ by
setting $A_\alpha = Z_{\alpha y}$ and $B_\alpha = Z_{\alpha x}$ in
(\ref{eq:ppale}).  Translational invariance of $\S$ is ensured by the
last equation in (\ref{eq:mini}).  Furthermore, not all $A_\alpha$ and
$B_\alpha$ can vanish in this case, since we assumed $\mat{Z}_\alpha
\neq \mat{Z}_\beta$ for some $\alpha \neq \beta$.

The residual angular momentum in a general linear gauge
(\ref{eq:ppale}) is given in the classical theory, to first order in
$\mat{\delta R}_\alpha$, by
\begin{equation}
  \label{eq:intrinsic}
  \begin{split}
    \Lambda & = \ver{z} \cdot \sum_{\alpha=1}^N m_\alpha
    \mat{Z}_\alpha \wedge \mat{\delta} \dot{\mat{R}}_\alpha + \ver{z}
    \cdot \sum_{\alpha=1}^N m_\alpha \mat{\delta R}_\alpha \wedge
    \mat{\delta} \dot{\mat{R}}_\alpha - \xi \left(
      \R^2 + 2 \sum_{\alpha=1}^N m_\alpha
      \mat{Z}_\alpha \cdot \mat{\delta R}_\alpha  \right) \\
    &\mbox{ } + \xi \frac{\Q(\{\mat{Z}_\alpha\})}{\R^2}
    \left(\Q(\{\mat{Z}_\alpha\}) + 2 \Q(\{\mat{\delta R}_\alpha\})
      \rule{0ex}{3ex}\right),
  \end{split}
\end{equation}
with $\R^2 = \sum_{\beta=1}^N m_\beta \mat{Z}_{\beta}^2$ and, 
\begin{equation}
  \label{eq:ecxi}
  \begin{aligned}
  \xi & = \frac{1}{\sum_{\beta=1}^N m_\beta \mat{R}_\beta^2} \left(
  \ver{z} \cdot \sum_{\alpha=1}^N m_\alpha \mat{R}_\alpha
  \wedge \dot{\mat{R}}_\alpha - \ell_z \right)
  = \frac{1}{\R^2}  \left( \rule{0ex}{5ex} 1 -
    \frac{2}{\R^2} \sum_{\beta=1}^N m_\beta \mat{Z}_\beta \cdot
  \mat{\delta R}_\beta \right) \times \\
  &  \times \left( \ver{z} \cdot \sum_{\alpha=1}^N
  m_\alpha 
  \mat{Z}_\alpha \wedge \mat{\delta} \dot{\mat{R}}_\alpha - \ell_z
  \right) + 
  \frac{1}{\R^2}
  \ver{z} \cdot \sum_{\alpha=1}^N m_\alpha \mat{\delta R}_\alpha
  \wedge \mat{\delta} \dot{\mat{R}}_\alpha + \mathcal{O}(\mat{\delta 
  R}_\alpha^2).   
  \end{aligned}
\end{equation}
The expression for $\Lambda$ in the linearized principal axes gauge is
obtained by substituting $A_\alpha= Z_{\alpha y}$ and $B_\alpha =
Z_{\alpha x}$ in (\ref{eq:intrinsic}).  In general, $\Lambda$ does not
vanish at the equilibrium positions $\mat{\delta R}_\alpha = 0$.  In
order to make $\Lambda$ of first order in $\mat{\delta R}_\alpha$ we
impose instead the gauge condition
\begin{equation}
  \label{eq:eckart}
  \sum_{\alpha=1}^N m_\alpha \mat{Z}_\alpha \wedge \mat{R}_\alpha = 
  \sum_{\alpha=1}^N m_\alpha \mat{Z}_\alpha \wedge \mat{\delta
  R}_\alpha = 0, 
\end{equation}
which corresponds to (\ref{eq:ppale}) with $A_\alpha= -Z_{\alpha y}$
and $B_\alpha = Z_{\alpha x}$.  The gauge condition (\ref{eq:eckart})
defines the Eckart frame \cite{ek2}.  The general expression
(\ref{eq:intrinsic}) for $\Lambda$, in this gauge simplifies to,
\begin{equation}
  \label{eq:intreckart}
  \Lambda  =  \ver{z} \cdot \sum_{\alpha=1}^N m_\alpha \mat{\delta
  R}_\alpha \wedge \mat{\delta} \dot{\mat{R}}_\alpha. 
\end{equation}
Fixing a gauge in which $\Lambda$ is of first order in $\mat{\delta
  R}_\alpha$ as in (\ref{eq:intreckart}) is a necessary condition to
satisfying Casimir's criterion for the decoupling of rotational and
vibrational degrees of freedom in a quasirigid system in low orders in
perturbation theory \cite{ek2}.  As shown above, such condition is not
fulfilled by the principal axes frame \cite{ek2}.

In Eckart gauge the momentum operators $\mat{\varPi}_\alpha$ conjugate
to $\mat{\delta R}_\alpha$ are given by (\ref{eq:qmom1}), with the
values of $A_\alpha$, $B_\alpha$ corresponding to (\ref{eq:eckart}),
and with derivatives $\partial / \partial X_\alpha$, $\partial /
\partial Y_\alpha$ substituted by $\partial / \partial \delta
X_\alpha$ and $\partial / \partial \delta Y_\alpha$.  The operator
$\Lambda$ is then of the form,
\begin{equation}
  \label{eq:intru}
  \Lambda = \frac{1}{i} \sum_{\alpha=1}^N \left( \delta X_\alpha
  \frac{\partial}{\partial \delta Y_\alpha} - \delta Y_\alpha
  \frac{\partial}{\partial \delta X_\alpha} \right) - \frac{1}{i}
  \frac{\Q(\{\mat{\delta R}_\gamma\})}{\R^2} \sum_{\alpha=1}^N \left(
  Z_{X\alpha} \frac{\partial}{\partial \delta Y_\alpha} - Z_{Y\alpha}
  \frac{\partial}{\partial \delta X_\alpha} \right).
\end{equation}
Its coefficients are of first order in $\mat{\delta R}_\alpha$.  The
Hamiltonian operator is obtained by substituting these expressions for
$\mat{\varPi}_\alpha$ and $\Lambda$, into (\ref{eq:qham}), with
$\Q(\{\mat{R}_\alpha\}) = \Q(\{\mat{Z}_\alpha\}) + \Q(\{\mat{\delta
  R}_\alpha\}) = \R^2 + \sum_{\gamma = 1}^N m_\gamma \mat{Z}_\gamma
\cdot \mat{\delta R}_\gamma$.  The inner product, finally, is given by
(\ref{eq:inp1}).

\subsection{The case \mat{N=2}}
\label{sec:N=2}

As a very minimal verification of the formalism we consider a system
of two particles interacting through an elastic potential that models
a spring with rest length $a$, $\V = k/2 (|\mat{r}_1 - \mat{r}_2| -
a)^2$.  We verify that the Hamiltonian operator of the previous
section, with the corresponding inner product, leads to the correct
energy spectrum and wave functions in a semiclassical expansion for
large $a$.  Since the residual angular momentum $\Lambda$ vanishes in
a translationally invariant system for $N < 3$, this simple example
does not provide an illustration of the role of terms linear in
$\ell_z$ in $\H$, nor of Casimir's condition.  In this section we
restore $\hbar$ in all expressions.

We choose the minimum $\mat{Z}_{1,2}$ of the potential as,
$(Z_{1,2})_x 
= \pm a m_{2,1}/M$, $(Z_{1,2})_y=0$, with $M=m_1+m_2$.  These
$\mat{Z}_{1,2}$ satisfy (\ref{eq:mini}).  The gauge conditions
defining the Eckart frame are then,
\begin{equation}
  \label{eq:2gau}
  \S(\{\mat{\delta R}_\alpha\}) = a \mu (\delta Y_1 - \delta Y_2)=0,
  \quad
  \mat{\C}(\{\mat{\delta R}_\alpha\}) = \frac{m_1}{M} \mat{\delta R}_1
  + \frac{m_2}{M} \mat{\delta R}_2 = 0,
\end{equation}
where $\mu$ is the reduced mass.  Together, (\ref{eq:2gau}) imply
$\delta Y_1 = 0 = \delta Y_2$.  The Faddeev-Popov determinant \Q\ and
$\R^2$ are given by,
\begin{equation}
  \label{eq:2fpd}
  \Q = \mu a (a + \delta X_1 - \delta X_2),
  \quad
  \R^2 = \mu a^2.
\end{equation}
The momentum operators (\ref{eq:qmom1}) are then,
\begin{equation}
  \label{eq:2qmom1}
  \varPi_{X_1} = \frac{\hbar}{i} \left( \frac{m_2}{M}
  \frac{\partial}{\partial \delta X_1} - \frac{m_1}{M}
  \frac{\partial}{\partial \delta X_1} \right) = - \varPi_{X_2},
  \quad
  \varPi_{Y_1} = 0 = \varPi_{Y_2},
\end{equation}
consistent with the gauge conditions (\ref{eq:2gau}).  In terms of
these operators we write the Hamiltonian (\ref{eq:nham}) as,
\begin{equation}
  \label{eq:2ham}
  \widetilde \H = \frac{1}{2 m_1} \varPi_{X_1}^2 + \frac{1}{2 m_2}
  \varPi_{X_2}^2 + \frac{\mu \omega^2}{2} (\delta X_1 - \delta X_2)^2
  + \frac{\hbar^2 \R^2}{2 \Q^2} \left(\ell_z^2 - \frac{1}{4}\right),
\end{equation}
where $k=\mu \omega^2$, and the last term gathers the centrifugal and
quantum potentials.  Using $\varPi_{X_1} = -\varPi_{X_2}$, we rewrite
$\widetilde \H$ as,
\begin{equation}
  \label{eq:2ham2}
  \begin{gathered}
    \widetilde \H = \widetilde \H_0 + \widetilde \H_1,
    \quad
    \widetilde \H_0 = \frac{1}{2\mu} \varPi_{X_1}^2 + \frac{\mu
      \omega^2}{2} (\delta X_1 - \delta X_2)^2,\\
    \widetilde \H_1 = \frac{\hbar^2 \R^2}{2 \Q^2} \left(\ell_z^2 -
    \frac{1}{4}\right) = \frac{\hbar \omega}{2} \epsilon^2 \left(\ell_z^2 -
    \frac{1}{4}\right) \left( 1 - 2 \epsilon \sqrt{\frac{\mu
    \omega}{\hbar}} (\delta X_1 - \delta X_2) + \mathcal{O}
    (\epsilon^2) \right).
  \end{gathered}
\end{equation}
In $\widetilde{\H_1}$ we denoted $\epsilon = \sqrt{\hbar/(\mu \omega
  a^2)}$, which is our perturbation expansion parameter.  Taking into
account (\ref{eq:2gau}), (\ref{eq:2fpd}) and the fact that the
Jacobian has been absorbed in the wave functions, the inner product
(\ref{eq:inp1}) takes the form,
\begin{equation}
  \label{eq:2inn}
    \langle \widetilde\phi | \widetilde\psi \rangle  = 
    \frac{1}{\mu a} \iint_{-\infty}^{+\infty} 
    d\delta X_1 d\delta
    X_2 \, \delta \left( \frac{m_1}{M} \delta X_1 + \frac{m_2}{M}
    \delta X_2 \right) \Theta(a+\delta X_1-\delta X_2)
    (\widetilde\phi^* \widetilde \psi)(X_1,0,X_2,0).
\end{equation}
Using (\ref{eq:2ham2}) and (\ref{eq:2inn}) we can compute the
perturbative expansion.  Since $[m_1 \delta X_1 + m_2 \delta X_2,
\varPi_{X_1}] = 0$, the eigenfunctions of $\widetilde\H_0$ depend on
$\delta X_{1,2}$ only through $\delta X_1 - \delta X_2$.  The
eigenvalues $E_{(0)n}$ and eigenfuntions $\phi_{(0)n}(\delta X_1 -
\delta X_2)$ of $\widetilde\H_0$ are then those of a one-dimensional
harmonic oscillator.

To order $\mathcal{O}(\epsilon^2)$, the perturbed energies can be read
off the expression (\ref{eq:2ham2}) for $\widetilde\H_1$, 
\begin{equation}
  \label{eq:2eng}
  E_n = E_{(0)n} + E_{(1)n},
  \qquad
  E_{(1)n} = \frac{\hbar \omega}{2} \epsilon^2 \left( \ell_z^2 -
  \frac{1}{4}\right). 
\end{equation}
Using the inner product (\ref{eq:2inn}), the perturbed wave functions
are found to be, to $\mathcal{O}(\epsilon^3)$,
\begin{equation}
  \label{eq:2wfct}
  \phi_n (\delta X_1 -\delta X_2) = \phi_{(0)n} + \frac{\epsilon^3}{2}
  \left( \ell_z^2 - \frac{1}{4}\right) \left( \sqrt{\frac{n}{2}}
  \phi_{(0)n-1} -  \sqrt{\frac{n+1}{2}} \phi_{(0)n+1}\right),
\end{equation}
where on the r.h.s.\ we omited the argument $(\delta X_1 -\delta X_2)$
of wave functions for brevity.  These results, (\ref{eq:2eng}) and
(\ref{eq:2wfct}), agree with a conventional perturbative calculation
as they should.

\section{Final remarks}
\label{sec:final}

We considered above the classical and quantum dynamics of
two-dimensional many-body systems in rotating reference frames, in a
gauge invariant approach \cite{als,bs1}.  Our treatment parallels the
formulation of gauge field theories in non-covariant gauges in the
Schr\"odinger representation \cite{lee,chr,kiw}, and generalizes to
$N$-body systems the analysis of the one-particle case of
\cite{lee,chr,bou}.

The gauge-invariant approach allows us to deal with constrained
degrees of freedom without necessarily solving the constraints.  Yet,
that approach entails also a reduction of configuration space by the
elimination of the angular degree of freedom.  The remaining,
constrained dynamical variables only span the reduced configuration
space.  Our formalism has a direct physical and geometrical
interpretation which we have tried to emphasize.  In the gauges of
sections \ref{sec:linear}--\ref{sec:eckart} it is not difficult to
obtain the metric tensor on the gauge hypersurface in terms of
coordinates $(\{\mat{R}_\alpha\},\theta)$ (and \mat{\rho}\ if there is
translation invariance).  In those coordinates the metric has as many
zero eigenvalues as the dimension of the gauge group, its kernel
comprising the subspace orthogonal to the gauge surface.  From the
restriction of that metric tensor to the gauge surface the kinetic
energy operator in the quantum theory can be constructed by applying
the usual expression for the Laplacian in curvilinear coordinates.

In sections \ref{sec:linear}--\ref{sec:cmm} we gave a derivation of
the quantum theory in rotating frames.  A detailed discussion is
provided there of Gribov ambiguities and of the commutator algebra,
which are essential for the obtention of the inner product and of the
momentum and Hamiltonian operators.  Those issues seem to us to have
been neglected in the previous literature.  Also discussed in detail
is the residual angular momentum operator $\Lambda$.  In linear gauges
its eigenvalues are integer.  In the quadratic gauge of section
\ref{sec:quadratic} its eigenvalues depend on the dynamical variables
$\{\mat{R}_\alpha\}$ though, remarkably, only through $Q$ and $R^2$
(see (\ref{eq:bbb}), (\ref{eq:omega}) and (\ref{eq:qquan})), which are
measures of the instantaneous shape of the system.  The relation
between wave functions in the laboratory frame and in rotating ones,
which we treat rather briefly, is summarized in the expression for the
inner product in the different gauges, (\ref{eq:xi0}), (\ref{eq:inp}),
(\ref{eq:cinp}) and (\ref{eq:inp1}).  Such relation is best understood
by going through the succesive steps of the derivation of the inner
product by the Faddeev-Popov technique.  For systems of identical
particles the formalism is clearly symmetric under permutations of
particles if the gauge conditions are chosen so that they are
symmetric in the position vectors $\{\mat{R}_\alpha\}$.

In the case of quasi-rigid systems we recovered in section
\ref{sec:eckart} some of Eckart's classic results \cite{ek2}.  In
particular, in the classical theory the residual angular momentum
$\Lambda$ becomes manifestly small of $\mathcal{O}(\mat{\delta R})$ in
Eckart gauge.  Thus, in a perturbative expansion the term $L_z
\Lambda$ in \H\ is of higher order than the term $L_z^2$.  This leads
to a decoupling of rotational ($L_z$) and vibrational ($\{\mat{\delta
  R}_\alpha\}$) degrees of freedom in low orders in perturbation
theory.  (In the elementary example of section \ref{sec:N=2}, in which
$\Lambda = 0$, the decoupling is apparent through
$\mathcal{O}(\epsilon^2)$ in (\ref{eq:2ham2}).)  Such decoupling would
not be manifest in other gauges in which $\Lambda \sim \mathcal{O}(1)$
\cite{ek2,ek1}, since in that case the terms $L_z^2$ and $L_z \Lambda$
would be of the same perturbative order.  That is the case of the
instantaneous principal axes frame \cite{ek2,ek1}.

We do not agree with the point of view of \cite{vil} (see also
\cite{bs1,bs2}) that, by quantizing a many-body system in the
instantaneous principal axes frame (see eq.\ (II.4) in \cite{vil}), it
is possible to separate the collective rotation from the intrinsic
dynamics.  If that statement were generally valid, in the quasi-rigid
case it should imply the decoupling of the total angular momentum from
the vibrational degrees of freedom in perturbation theory.  But, as
shown in \cite{ek2}, that decoupling is not manifest in the principal
axes frame.

\section*{Acknowledgements}

We would like to thank C.\ A.\ Garc\'{\i}a Canal and V.\ Gupta for
their critical comments to a preliminary version of this paper.  This
work has been partially supported by Conacyt of Mexico through grant
32598E.

\appendix
\numberwithin{equation}{section}

\section{Electrodynamics}
\label{sec:appqed}

The methods used in the preceding sections are common to all (Abelian)
gauge theories and, therefore, have a direct counterpart in
electrodynamics.  In this appendix we briefly discuss those aspects of
electrodynamics, following the treatment of \cite{chr,lee,kiw}.  In
terms of the field-strength tensor $F^{\mu\nu} = \partial^\mu A^\nu -
\partial^\nu A^\mu$, the Lagrangian density of the e.m.\ field coupled
to an external current density $j^\mu$ has the familiar expression,
\begin{equation}
  \label{eq:em1}
  \Ll = -\frac{1}{4} F^{\mu\nu} F_{\mu\nu} - j^\mu A_\mu.
\end{equation}
We assume that the current is conserved, $\partial^\mu j_\mu = 0$, so
that the action of the system is invariant under the e.m.\ gauge
transformations $A^{\mu\prime} = A^\mu - \partial^\mu \Lambda$, with
$\Lambda = \Lambda(x)$ an arbitrary function of the space-time
coordinates.  We consider only fields, currents, and gauge-parameter
functions vanishing at spatial infinity.  Furthermore, we assume that
$j^\mu$ is such that there is an inertial frame in which $\partial^0
j^0 = 0$, and therefore $\mat{\nabla}\cdot \mat{j} =0$.  We choose
that reference frame to formulate the Hamiltonian formalism.  The
extension to the general case $\partial^0 j^0 \neq 0$, which involves
time-dependent constraints, is not necessary for our purposes.

\paragraph{Weyl gauge}

Since the time derivative of $A^0$ does not enter $\Ll$, it is an
auxiliary field.  By means of a gauge transformation any field
configuration can be brought into the form $A^0=0$.  We denote fields
in Weyl gauge by $V^\mu (x)$, with $V^0 = 0$.  In this gauge the
Lagrangian density reduces to,
\begin{equation}
  \label{eq:em2}
  \Ll = \frac{1}{2} \dot{\mat{V}}^2 - \frac{1}{2} (\mat{\nabla} \wedge
  \mat{V})^2 + \mat{j} \cdot \mat{V}.
\end{equation}
$\Ll$ must be supplemented by the equation of motion for $V^0$ derived
from (\ref{eq:em1}), which is Gauss law,
\begin{equation}
  \label{eq:em3}
  \mat{\nabla} \cdot \dot{\mat{V}} = - j^0,
\end{equation}
with $\dot{\mat{V}} = -\mat{E}$.  In the Hamiltonian formulation eq.\ 
(\ref{eq:em3}) is written in terms of the momenta
canonically conjugate to \mat{V}, $-\mat{E}$, and constitutes a
primary first-class constraint.  It does not give rise to further
secondary constraints.  The Hamiltonian density is given by,
\begin{equation}
  \label{eq:em4}
  \H = \frac{1}{2} \mat{E}^2 + \frac{1}{2} (\mat{\nabla}\wedge
  \mat{V})^2 - \mat{j} \cdot \mat{V},
  \quad
  \mat{\nabla} \cdot \mat{E} = j^0.
\end{equation}
The Poisson brackets are canonical,
\begin{equation}
  \label{eq:em5}
  \left[ E^j(t,\mat{x}), V^k(t,\mat{y})  \right]_P = \delta^{jk}
  \delta(\mat{x}-\mat{y}). 
\end{equation}
All other brackets among basic dynamical variables vanish.  The theory
in this gauge is invariant under gauge transformations with a
time-independent parameter $\Lambda(\mat{x})$.  The canonical
generator of that symmetry is the l.h.s.\ of Gauss law, $\mat{\nabla}
\cdot \mat{E} - j^0$ (or just $\mat{\nabla} \cdot \mat{E}$).  We have
the Poisson brackets,
\begin{equation}
  \label{eq:em6}
  \left[ \mat{\nabla} \cdot \mat{E}(t,\mat{x}) - j^0(\mat{x}), H
  \right]_P = 0, 
  \quad
  \left[ \mat{\nabla} \cdot \mat{E}(t,\mat{x}) - j^0(\mat{x}),
  V^k(t,\mat{y}) \right ]_P = \partial^k \delta(\mat{x} - \mat{y}), 
\end{equation}
where $H=\int d^3\mat{x} \H(t,\mat{x})$.  The first eq.\ in
(\ref{eq:em6}) expresses the consistency of the constraint with the
dynamics, and the second shows that the constraint is the infinitesimal
generator of the residual gauge symmetry.  

In the quantum theory in the Schr\"odinger representation the field
operators are time-independent, the basic commutators are obtained
from (\ref{eq:em5}), and the Hamiltonian density operator is given by,
\begin{equation}
  \label{eq:em6a}
  \H (\mat{x}) = -\frac{1}{2} \frac{\delta^2}{\delta V^k(\mat{x})
  \delta V^k(\mat{x})} + \frac{1}{2} (\mat{\nabla} \wedge
  \mat{V}(\mat{x}))^2 - \mat{j}(\mat{x}) \cdot
  \mat{V}(\mat{x}). 
\end{equation}
Gauss law is imposed as a constraint on the state space of the
system,
\begin{equation}
  \label{eq:em6b}
  \frac{1}{i} \mat{\nabla} \cdot \frac{\delta}{\delta
  \mat{V}(\mat{x})} 
  \Psi[\mat{V}] = j^0(\mat{x}) \Psi[\mat{V}],
\end{equation}
where $\Psi[\mat{V}]$ is the wave functional.

\paragraph{Coulomb gauge}

Any field $V^\mu$ in Weyl gauge can be transformed into Coulomb gauge,
$\mat{\nabla} \cdot \mat{A} = 0$, and conversely,
\begin{subequations}
\begin{gather}
  \label{eq:em7a}
  A^\mu(t,\mat{x}) = V^\mu(t,\mat{x}) + \partial^\mu
  \Lambda(t,\mat{x}), \quad \Lambda(t,\mat{x}) =
  \frac{1}{\mat{\nabla}^2} \left( \mat{\nabla} \cdot
    \mat{V}(t,\mat{x}) \rule{0ex}{2ex} \right) \equiv -\frac{1}{4 \pi}
  \int 
  d^3\mat{x}^\prime \frac{\mat{\nabla}^\prime \cdot
    \mat{V}(t,\mat{x}^\prime)}{|\mat{x}^\prime - \mat{x}|},\\
  \label{eq:em7b}
  V^\mu(t,\mat{x}) = A^\mu(t,\mat{x}) - \partial^\mu
  \lambda(t,\mat{x}), \quad \lambda(t,\mat{x}) = \int_{t_0}^t
  dt^\prime A^0(t^\prime,\mat{x}).
\end{gather}
\end{subequations}
$A^0$ in (\ref{eq:em1}) is an auxiliary field which is determined by
its Lagrangian equation, 
\begin{equation}
  \label{eq:em8}
  A^0 = -\frac{1}{\mat{\nabla}^2} j^0.
\end{equation}
Once the gauge has been fixed, we can substitute this expression back
into the Lagrangian, to obtain, 
\begin{equation}
  \label{eq:em9}
  \Ll = \frac{1}{2} \dot{\mat{A}}^2 - \frac{1}{2} (\mat{\nabla} \wedge 
  \mat{A})^2 + \frac{1}{2} j^0 \frac{1}{\mat{\nabla}^2} j^0 + \mat{j}
  \cdot \mat{A},
  \quad
  \mat{\nabla} \cdot \mat{A} = 0.
\end{equation}
The electric field is 
\begin{equation}
  \label{eq:em9+}
  \mat{E} = \mat{E}_T + \mat{E}_L,
  \quad
  \mat{E}_L = -\mat{\nabla} A^0 = \mat{\nabla}
  \frac{1}{\mat{\nabla}^2} \mat{\nabla}\cdot \mat{E}, 
  \quad
  \mat{E}_T = -\dot{\mat{A}}.
\end{equation}
The momentum conjugate to \mat{A}\ is $\mat{\varPi} = -\mat{E}_T$.
Substituting (\ref{eq:em7b}) and (\ref{eq:em9+}) into (\ref{eq:em4}),
we get the classical Hamiltonian density in this gauge,
\begin{equation}
  \label{eq:em10}
  \H = \frac{1}{2} \mat{\varPi}^2 + \frac{1}{2} (\mat{\nabla} \wedge
  \mat{A})^2 - \frac{1}{2} j^0 \frac{1}{\mat{\nabla}^2} j^0 - \mat{j}
  \cdot \mat{A},
  \quad
  \mat{\nabla} \cdot \mat{A} = 0 = \mat{\nabla} \cdot \mat{\varPi}.
\end{equation}
Furthermore, from the inverse transformation (\ref{eq:em7a}) and
(\ref{eq:em9+}), and the Poisson brackets (\ref{eq:em6}), we can
compute the brackets among coordinates and momenta in this gauge.  The
corresponding equal-time quantum commutators are,
\begin{equation}
  \label{eq:em12}
  \left[ A^j(t,\mat{x}), \mat{E}^k_T(t,\mat{x}^\prime) \right] =
  i \left( -\delta^{jk} + \nabla^j \nabla^{\prime k}
  \frac{1}{\mat{\nabla}^2}\right) 
  \delta(\mat{x} - \mat{x}^\prime).
\end{equation}
The gauge condition $\mat{\nabla} \cdot \mat{A} =0$ and the derived
relation $\mat{\nabla} \cdot \mat{\varPi} =0$ constitute second-class 
constraints, valid as operator equations.  Their associated Dirac
brackets are given by (\ref{eq:em12}). 

A realization of the commutator algebra (\ref{eq:em12}) in the
Schr\"odinger representation can be found as in sections
\ref{sec:linear} and \ref{sec:quadratic},
\begin{equation}
  \label{eq:em14}
  \varPi^j(\mat{x}) = \frac{1}{i} \frac{\delta}{\delta
  A^j(\mat{x})} - \frac{1}{i} \nabla^j \nabla^k
  \frac{1}{\mat{\nabla}^2} \frac{\delta}{\delta A^k(\mat{x})}, 
\end{equation}
where the repeated index $k$ is summed from 1 to 3.  We can express
the momentum operators in Weyl gauge in terms of the operators
(\ref{eq:em14}) by means of the chain rule,
\begin{equation}
  \label{eq:em15}
  \frac{\delta}{\delta V^m(\mat{x})} = \int d^3\mat{y} \frac{\delta
  A^k(\mat{y})}{\delta V^m(\mat{x})} \frac{\delta}{\delta
  A^k(\mat{y})}.  
\end{equation}
From (\ref{eq:em7a}) we see that,
\begin{equation}
  \label{eq:em16}
  \frac{\delta
  A^k(\mat{x}^\prime)}{\delta V^m(\mat{x})} = \delta^{km} \delta
  (\mat{x} - \mat{x}^\prime) - \frac{1}{4\pi} \nabla^{\prime k}
  \nabla^m \left( \frac{1}{|\mat{x}^\prime - \mat{x}|}\right). 
\end{equation}
Replacing (\ref{eq:em16}) into (\ref{eq:em15}), we get,
\begin{equation}
  \label{eq:em17}
  \frac{\delta}{\delta V^m(\mat{x})} = i \varPi^m(\mat{x}),
\end{equation}
and therefore, from the Hamiltonian density operator (\ref{eq:em6a})
in Weyl gauge, we obtain the operator in Coulomb gauge which has the
same form as the classical density (\ref{eq:em10}) with \mat{\varPi}\
from (\ref{eq:em14}).  The Faddeev-Popov Jacobian in this case is
field-independent, as can be seen from (\ref{eq:em16}), which is why
it does not appear in the Hamiltonian.

\section{Eigenvalues and eigenfunctions of $\Lambda$}
\label{sec:appL}

In this appendix we discuss the eigenvalues and eigenfunctions of the
residual angular momentum $\Lambda = \sum_{\beta=1}^N \left( X_\beta
  \varPi_{Y_\beta} - Y_\beta \varPi_{X_\beta} \right)$.  Since the
definition of $\mat{\varPi}_\alpha$ depends on the gauge conditions,
the form of $\Lambda$ as a differential operator is different in the
cases of linear and quadratic gauge conditions.

\subsection{Linear gauges}
\label{sec:applin}

In the linear gauges of sections \ref{sec:linear}, \ref{sec:cmm} and
\ref{sec:eckart} $\Lambda$ is given by (\ref{eq:intri}).  In order to
find its eigenvalues and eigenfunctions we apply the method of
characteristic lines \cite{wbr} (which can also be applied, of course,
to the standard angular momentum operator in Cartesian coordinates).
It is therefore necessary to obtain first the classical orbits
generated by $\Lambda$.

\paragraph{Classical orbits}

The orbits generated by the classical $\Lambda$ on the gauge surface
in configuration space are described by the equations,
\begin{equation}
  \label{eq:dorb}
  \frac{dX_\gamma}{d\alpha} = [\Lambda, X_\gamma]_P,
  \qquad
  \frac{dY_\gamma}{d\alpha} = [\Lambda, Y_\gamma]_P,
  \qquad
  \S(\{\mat{R}_\gamma\}) = 0.
\end{equation}
With the Poisson brackets given in (\ref{eq:+cmm}) (as quantum
commutators), we obtain the solution to (\ref{eq:dorb}) as,
\begin{equation}
  \label{eq:orbits}
  \begin{aligned}
    X_\gamma(\alpha) & = \exp \left([\Lambda,\bullet]_P
    \rule{0ex}{2ex} \right)
    X_\gamma = \frac{B_\gamma \Q}{\R^2} + \cos \alpha \left(X_\gamma -
      \frac{B_\gamma \Q}{\R^2} \right) + \sin \alpha \left(Y_\gamma +
      \frac{A_\gamma \Q}{\R^2} \right), \\
    Y_\gamma(\alpha) & = \exp \left([\Lambda,\bullet]_P
    \rule{0ex}{2ex} \right) Y_\gamma =
    -\frac{A_\gamma \Q}{\R^2} + \cos \alpha \left(Y_\gamma +
      \frac{A_\gamma \Q}{\R^2} \right) - \sin \alpha \left(X_\gamma -
      \frac{B_\gamma \Q}{\R^2} \right).
  \end{aligned}
\end{equation}
If $\S(\{\mat{R}_\gamma(0)\}) = 0$, then
$\S(\{\mat{R}_\gamma(\alpha)\}) = 0$ for all $\alpha$.  In that case
\Q\ is constant along the orbits.  The center of mass also vanishes
for all $\alpha$ if it vanishes for $\alpha = 0$, as long as $\S$ is
translationally invariant, as explained in section \ref{sec:cmm}.

\paragraph{Kernel}

We consider first functions $C$ satisfying $\Lambda C =0$.  From
(\ref{eq:orbits}), we see that the quantities
\begin{equation}
  \label{eq:radii}
  \rho_\gamma^2 = \left( X_\gamma - \frac{B_\gamma \Q}{\R^2}
  \right)^2 + \left( Y_\gamma + \frac{A_\gamma \Q}{\R^2}  \right)^2
\end{equation}
are constant on the classical orbits.  Thus, any $f =
f(\{\rho_\beta\})$ satisfies $\Lambda f = 0$.  Furthermore, from
(\ref{eq:++cmm}) we have $\Lambda\Q = i \S$, so that for any function
$f = f(\Q)$ we get $\Lambda f(\Q) = f^\prime(\Q) \S$ which vanishes on
the gauge surface.  We therefore have $\Lambda C = 0$ if $C =
C(\{\rho_\beta\},\Q)$ and $\S = 0$. 

\paragraph{Eigenvalues and eigenfunctions}

Next, we consider the eigenvalue equation $i \Lambda \Psi = i \lambda
\Psi$, with $\lambda \neq 0$.  The characteristic lines of this
equation are defined by the differential system
\begin{equation}
  \label{eq:charlin}
  \frac{dX_1}{-\left(Y_1+\frac{A_1\Q}{\R^2}\right)} = \cdots =
  \frac{dX_N}{-\left(Y_N+\frac{A_N\Q}{\R^2}\right)} =
  \frac{dY_1}{X_1-\frac{B_1\Q}{\R^2}} = \cdots =
  \frac{dY_N}{X_N-\frac{B_N\Q}{\R^2}} =
  \frac{d\Psi}{i\lambda \Psi},
  \quad
  \S(\{\mat{R}_\gamma\}) = 0.
\end{equation}
Writing $\Psi = \exp(i\lambda \alpha)$, the solutions of
(\ref{eq:charlin}) are given by the classical orbits $\{X_\gamma
(-\alpha), Y_\gamma(-\alpha)\}$ from (\ref{eq:orbits}).  A solution
$\alpha$ to the eigenvalue equation is obtained by inverting a
relation of the form,
\begin{equation*}
  G(X_1(0),\ldots,X_N(0),Y_1(0),\ldots,Y_n(0)) = 0.
\end{equation*}
A possible choice is $G = Y_\gamma(0)+\frac{A_\gamma \Q}{\R^2}$.
Setting $Y_\gamma(\alpha) = Y_\gamma$ and $X_\gamma(\alpha) =
X_\gamma$, we are led to
\begin{equation*}
  \tan \alpha_\gamma = \frac{\R^2 Y_\gamma+\Q A_\gamma}{\R^2
  X_\gamma-\Q B_\gamma}. 
\end{equation*}
From here we obtain the family of eigenfuntions,
\begin{equation}\label{eq:psi}
  \Psi  = C \exp \left( i \sum_{\gamma =1}^N \lambda_\gamma
    \alpha_\gamma\right),
  \quad \mathrm{with} \quad
  \sum_{\gamma = 1}^N \lambda_\gamma = \lambda.
\end{equation}
The proportionality ``constant'' $C$ in (\ref{eq:psi}) can be any
function belonging to the kernel of $\Lambda$, as described above.
For $\Psi$ to be single-valued $\lambda_\gamma$, and therefore
$\lambda$, must be integers.

\subsection{Instantaneous principal axes gauge}
\label{sec:appppal}

In the quadratic gauge of section \ref{sec:quadratic} $\Lambda$ has
the form (\ref{eq:cint}).  We proceed as in the previous case,
starting from the classical orbits generated by $\Lambda$.  The
notation is the same as in section \ref{sec:quadratic}.

\paragraph{Classical orbits}

In this gauge we have the following equations for the classical
orbits on the gauge surface,
\begin{equation}
  \label{eq:qorb}
  \frac{dX_\gamma}{d\alpha} = [\Lambda, X_\gamma]_P =
  \left(1 + \frac{2Q}{R^2}\right) Y_\gamma,
  \quad
  \frac{dY_\gamma}{d\alpha} = [\Lambda, Y_\gamma]_P =
  - \left(1 - \frac{2Q}{R^2}\right) X_\gamma,
  \quad
  S(\{\mat{R}_\gamma\}) = 0.
\end{equation}
Their solution is,
\begin{equation}
  \label{eq:qorbits}
  \begin{aligned}
    X_\gamma (\alpha) & = \exp \left([\Lambda,\bullet]_P
      \rule{0ex}{2ex}\right) X_\gamma = \cos \left( \Omega \alpha
    \right) X_\gamma + \left( \frac{R^2+2Q}{R^2-2Q}\right)^{1/2} \sin
    \left(
      \Omega \alpha \right) Y_\gamma,\\
    Y_\gamma (\alpha) & = \exp \left([\Lambda,\bullet]_P
      \rule{0ex}{2ex} \right) Y_\gamma = -\left(
      \frac{R^2-2Q}{R^2+2Q}\right)^{1/2} \sin \left( \Omega \alpha
    \right) X_\gamma + \cos \left( \Omega \alpha \right) Y_\gamma,
  \end{aligned}
\end{equation}
with,
\begin{equation}
  \label{eq:omega}
  \Omega = \left(1 - \frac{4 Q^2}{R^4}\right)^{1/2}.
\end{equation}
The condition $S=0$ is preserved along these orbits and, if it is
satisfied, $Q$ and $R^2$ are constant on them.  Notice that we
require $Q \geq 0$ due to the Gribov ambiguity (see section
\ref{sec:quadratic}) and, therefore, by definition $0\leq 2 Q \leq
R^2$. 

\paragraph{Kernel}

The quantities
\begin{equation}
  \label{eq:rhos}
  \rho_\gamma^2 = \left( 1 - \frac{2 Q}{R^2} \right) X_\gamma^2 +
  \left( 1 + \frac{2 Q}{R^2} \right) Y_\gamma^2
\end{equation}
are constant along the orbits (\ref{eq:qorbits}).  Thus, if $S=0$, any
function $C = C(\{\rho_\gamma^2\}, Q, R^2)$ satisfies $\Lambda C = 0$.

\paragraph{Eigenvalues and eigenfunctions}

The characteristic lines of the equation $i\Lambda\Psi =
i\lambda\Psi$, $\lambda \neq 0$, in this gauge are the solutions to
the system,
\begin{equation}
  \label{eq:qchar}
  \frac{dX_1}{-Y_1 \left(1+\frac{2 Q}{R^2}\right)} = \cdots =
  \frac{dX_N}{-Y_N \left(1+\frac{2 Q}{R^2}\right)} = 
  \frac{dY_1}{X_1 \left(1-\frac{2 Q}{R^2}\right)} = \cdots =
  \frac{dY_N}{X_N \left(1-\frac{2 Q}{R^2}\right)} = 
  \frac{d\Psi}{i\lambda \Psi},
\end{equation}
together with the gauge condition $  S(\{\mat{R}_\gamma\}) = 0$.  As
in the previous section, we set $\Psi = \exp(i\lambda \alpha)$.  With
this, the solution to (\ref{eq:qchar}) is given by $\{X_\gamma
(-\alpha), Y_\gamma(-\alpha)\}$, see (\ref{eq:qorbits}).  Inverting
the relation $Y_\gamma(0) = 0$, we get,
\begin{equation}
  \label{eq:qualp}
  \alpha_\gamma = \frac{1}{\Omega} \mathrm{arctan}\left( \left(
    \frac{R^2+2 Q}{R^2 -2Q}\right)^{1/2}
    \frac{Y_\gamma}{X_\gamma}\right). 
\end{equation}
We then have,
\begin{equation}
  \label{eq:aaa}
  \Psi = C \exp\left( i \sum_{\gamma=1}^N \lambda_\gamma
    \alpha_\gamma\right),
  \quad
  \lambda = \sum_{\gamma=1}^N \lambda_\gamma.
\end{equation}
For $\Psi$ to be singled-valued we need,
\begin{equation}
  \label{eq:bbb}
  \lambda_\gamma = n_\gamma \Omega,
  \quad
  \lambda = n \Omega,
  \quad
  n = \sum_{\gamma=1}^N n_\gamma,
\end{equation}
with $n_\gamma$ integers.  We see that the eigenvalues $\lambda$ are
not integers in general, and that they depend through $\Omega$ on
$Q(\{\mat{R}_\alpha\})$ and $R^2(\{\mat{R}_\alpha\})$, which in turn
are functions of the dynamical variables $\{\mat{R}_\alpha\}$.
\end{document}